\documentclass[12pt,preprint]{aastex}

\usepackage{psfig}

\def\xmm{{\it XMM-Newton}}
\def\xte{{\it RXTE}}
\def\sax{{\it Beppo-SAX}}
\def\Nh{$N_{\rm H}$}

\slugcomment{to appear in ApJ 20 December 2004, v617}

\shorttitle{The Nucleus of NGC~6300 observed with \XMM}
\shortauthors{C. Matsumoto et al.}

\begin{document}

\title{An \xmm\  Observation of the Seyfert 2 Galaxy NGC~6300. \\
I. The Nucleus}

\author{Chiho Matsumoto\altaffilmark{1}, Aida Nava, Larry A. Maddox, Karen M. Leighly}
\affil{Department of Physics and Astronomy, 
	The University of Oklahoma,
    440 West Brooks Street, Norman, OK 73019}
\altaffiltext{1}{Current address: EcoTopia Science Institute, 
Nagoya University, Furo-cho, Chikusa, Nagoya, 464-8603, Japan;
Email: chiho@u.phys.nagoya-u.ac.jp}

\author{Dirk Grupe}
\affil{Astronomy Department, The Ohio State University, 
	140 West 18th Avenue, Columbus, OH 43210}

\author{Hisamitsu Awaki}
\affil{Department of Physics, Faculty of Science, Ehime University, 
Bunkyo-cho, Matsuyama, Ehime 790-8577, Japan}

\and

\author{Shiro Ueno}
\affil{Institute of Space and Astronautical Science,
      Japan Aerospace Exploration Agency,
      2-1-1 Sengen, Tsukuba, Ibaraki 305-8505, Japan
}

\begin{abstract}
We present results from a half-day observation by \xmm\ of the nucleus
of the nearby Seyfert 2 galaxy NGC~6300.  The X-ray spectrum of the
nucleus consists of a heavily absorbed hard component dominating 
the 3--10~keV band and a soft component seen in the 0.2--2~keV band.
In the hard band, the spectrum is well fitted by a power-law model with
photon index of 1.83$\pm$0.08 attenuated by a Compton-thin absorber 
($N_{\rm H}\simeq 2.2 \times 10^{23}\,\rm cm^{-2}$).  
A narrow iron line is detected at
6.43$_{-0.02}^{+0.01}$~keV with an equivalent width of $\sim$150~eV;
the line velocity width is marginally resolved to be
$\sigma\sim60$~eV.  The soft emission can be modeled as a power-law
and may be emission scattered by surrounding plasma.  Rapid and 
high-amplitude variability is observed in the hard X-ray band, whereas both
the iron line and the soft emission show no significant variability.
It is suggested that the nucleus has experienced an overall long-term
trend of decreasing hard X-ray intensity on a timescale of years.  We
discuss the origins of the spectral components.

\end{abstract}


\section{Introduction}

NGC 6300 is a nearby (z=0.0037; Mathewson \& Ford 1996) ringed, barred
spiral galaxy, classified as a SB(rs)b from its morphology, and also
identified as a bright Seyfert 2 from spectroscopy.  It was first
detected in hard X-rays in 1991 February during a {\it Ginga} maneuver
\citep{awakiD}.  Then, it was observed in the 3--24 keV band by \xte\
in 1997 February, and in the 0.1--200 keV band by \sax\ in 1999 August.
\xte\ measured a flat continuum spectrum (photon index: $\Gamma
\simeq 0.68$) with a superimposed K-$\alpha$ neutral iron emission
line of large equivalent width (EW$ \simeq 470$~eV).  These spectral
properties imply the presence of a Compton thick absorber obscuring
the nucleus \citep{lei99}.  Two-and-a-half years later, \sax\ obtained
a spectrum that was brighter in the whole 3--20 keV band, seen through
a Compton thin absorber ($N_{H} \simeq$ 2.1--3.1 $\times$10$^{23}\,\rm
cm^{-2}$), but with an iron line of the same intensity as the \xte\ line
\citep{gua02}. 
The spectral differences
between the two observations are most likely associated with the
transient behavior of the Seyfert nucleus, which was probably caught
in a high activity stage during the \sax\ observation \citep{gua02}. 
NGC~6300 is one of a few known
objects exhibiting transitions between Compton-thin and Compton-thick
X-ray spectral states \citep{mat03}.

In this paper we present the results from a new observation of
NGC~6300 with \xmm\ \citep{ref:xmm}, focusing on the spectral properties
and the resulting physical view of the nucleus.  Preliminary results
have been presented by \citet {aida} and \citet{lam02}.  Section 2
describes the observation and the data preparation, in \S3 we perform
image analyses in the soft and hard X-ray bands, and in \S4 we examine
variability in some energy bands.  Using the results from \S3 and
\S4, we perform a spectral analysis.  Finally, we discuss the physical
conditions of the spectral components and speculate on their origins.

Throughout this paper error bars in figures are 1\,$\sigma$,
and uncertainties quoted in the body and tables are 90\% 
confidence for one parameter of interest.

\section{Observation and Data Preparation}

NGC~6300 was observed from 2001 March 2 3:35 to 16:36 (UT) 
with \xmm.
The EPIC instruments consisting of one pn \citep{ref:pn} 
and two MOS \citep{ref:mos} CCDs
were operated in full-frame imaging mode using the medium filter.
The background during the whole observation
was low so that the entire observation could be used.
We did not analyze the Reflection Grating Spectrometer (RGS; den
Herder et al., 2001) data because of the low number of photons.

The data could not be reduced in the standard way because the
observation was split into two observation IDs (ObsIDs), 0059770101 between
03:35--05:32 and 0059770201 between 05:43--16:34.  The first
ObsID contained the pn and MOS data, while the second
contained the RGS and Optical Monitor (OM) data.  However, while the
event files in these ObsIDs were complete, the housekeeping
data only contained the times given above for the ObsIDs. 
Therefore, the housekeeping data had to be merged. We merge the
files using the {\sf FTOOLS} task {\sf fmerge} 
as described on the MPE Cookbook page.\footnote{
Available at http://wave.xray.mpe.mpg.de/xmm/cookbook/EPIC\_PN/merge\_odf.html.}
After this procedure the  EPIC event lists were created
in the standard way using the Science Analysis Software ({\sf SAS}) version 
5.3.0.
We filtered further using {\sf HEAsoft} versions 5.1 and 5.2.
We used {\sf SAS} version 5.3.3 to create the response 
matrices.\footnote{We had used the latest version when the analysis
was done. To check consistency with the newer calibration, we reduced the
data and created response matrices using the latest {\sf SAS} version
5.4.1. The parameters of our final model in Table~\ref{tab:wholespc} 
were consistent with those from {\sf SAS} 5.4.1. }
Throughout we used the ``flag=0''
events with the pattern of 0--4 and 0--12 from
the pn and MOS detectors, respectively.

\section{Image Analysis}

\begin{figure}
\plottwo{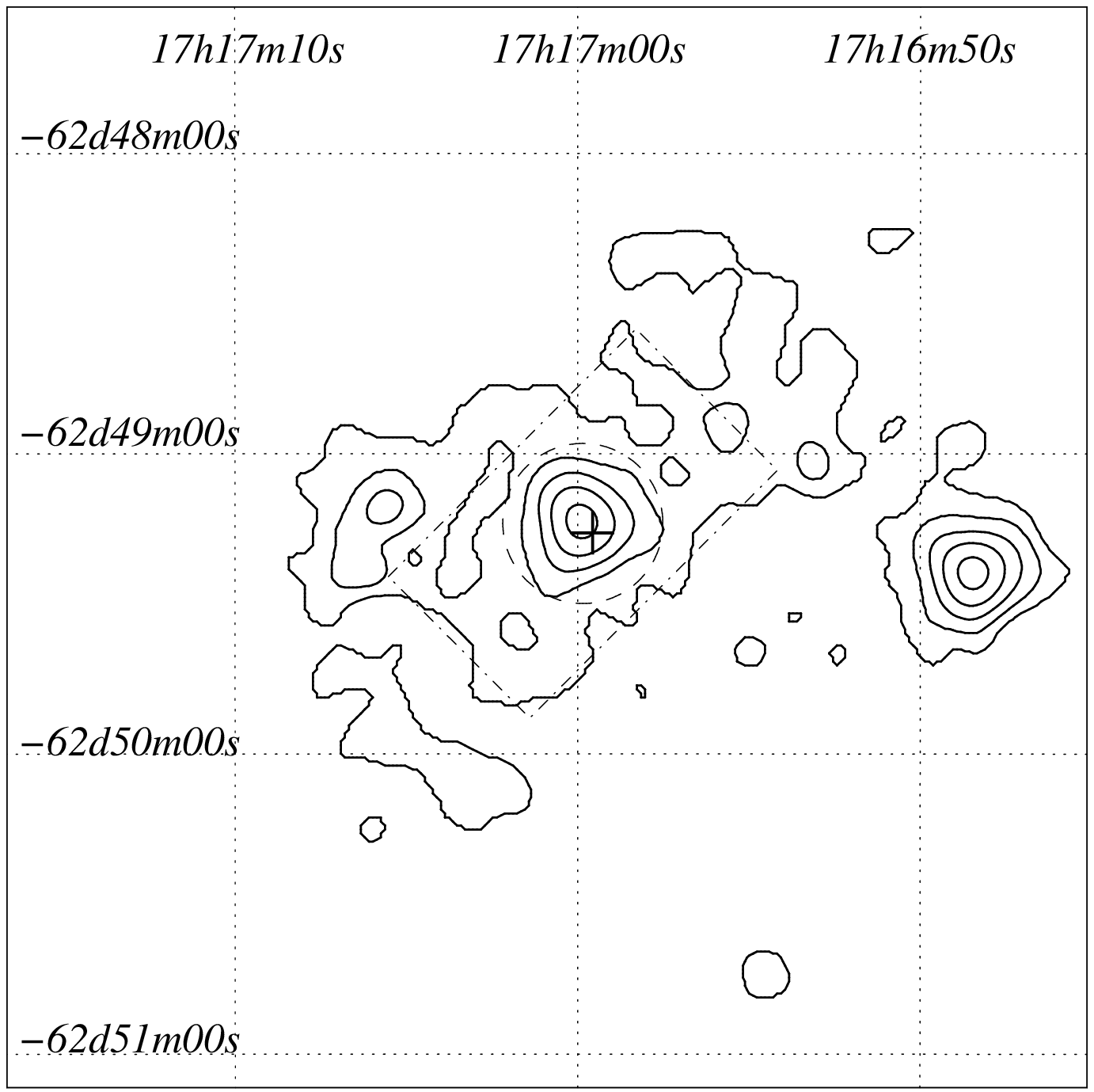}{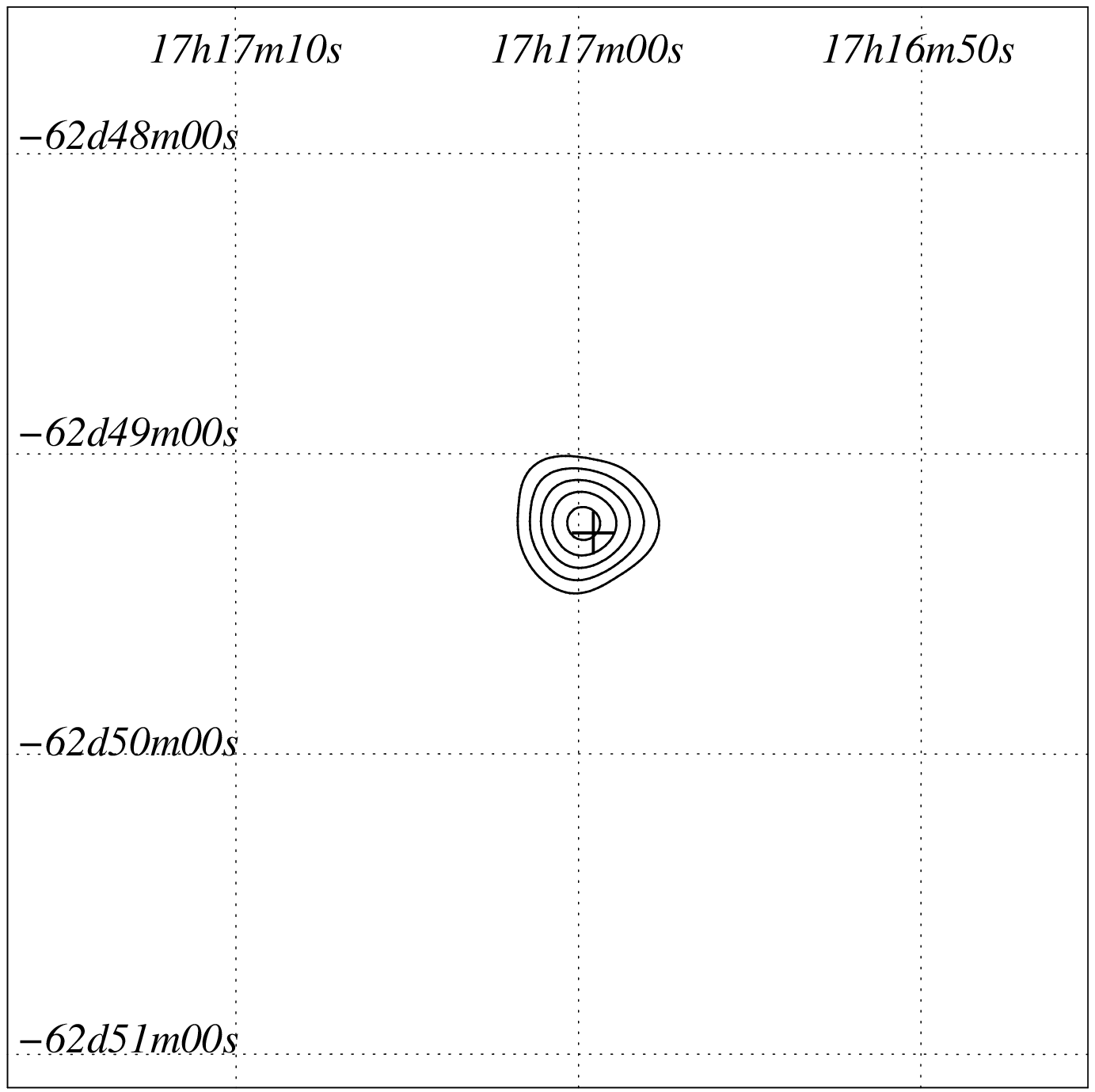}
\caption{EPIC images of NGC~6300 in the energy bands of 
0.2--2~keV (left) and 2--10~keV (right). 
Since the source falls near the chip gap of the pn, 
only the MOS data are used to make these plots. 
The images are smoothed with $\sigma=3.2^{''}$.
The contours are drawn 
such that each step corresponds to a factor of 1.8 difference, and
the first and last contours differ by an order of magnitude.
In the left figure, the extraction region ($r<16^{''}$) for the soft spectrum
is shown  by the dashed line.
A set of soft background spectra (BGD2; see \S\ref{sec:sft}) was
collected from the rectangular region enclosed by the dot-dashed line
excluding the $r<20^{''}$ from the nucleus.
The radio core position is marked by the cross.
}
\label{fig:img}
\end{figure}

Figure~\ref{fig:img} shows the X-ray contour plots in the soft and
hard X-ray bands.  In the hard X-ray image, the primary source is
located at R.A.$={\rm 17^h\, 16^m\, 59.8^s}$ and decl.$=-62^\circ\,
49'\, 13''$ (J2000.0), which is consistent with the position of the
radio (13~cm) core emission.\footnote{R.A.$={\rm 17^h\, 16^m\,
59.6^s}$, decl.$=-62^\circ\, 49'\, 16''$ with the beam size of
$\sigma\sim4^{''}$ (Maddox et al., in prep.)}  In the soft band, a
point-like source is detected at the position of the primary hard
X-ray source.  At least two more point-like sources are detected
within 80$^{''}$ separation from the nucleus, and emission from the
host galaxy is also detected around the nucleus.  One of the point
sources may be a new candidate ultra-luminous X-ray source.  Since the
origin of these emissions is difficult to determine solely from the
X-ray observation, details of the point-source properties and galaxy
emission will be presented in a separate paper involving analysis of
multi-wavelength observations (Maddox et al., in preparation).
 
\begin{figure}
\plotone{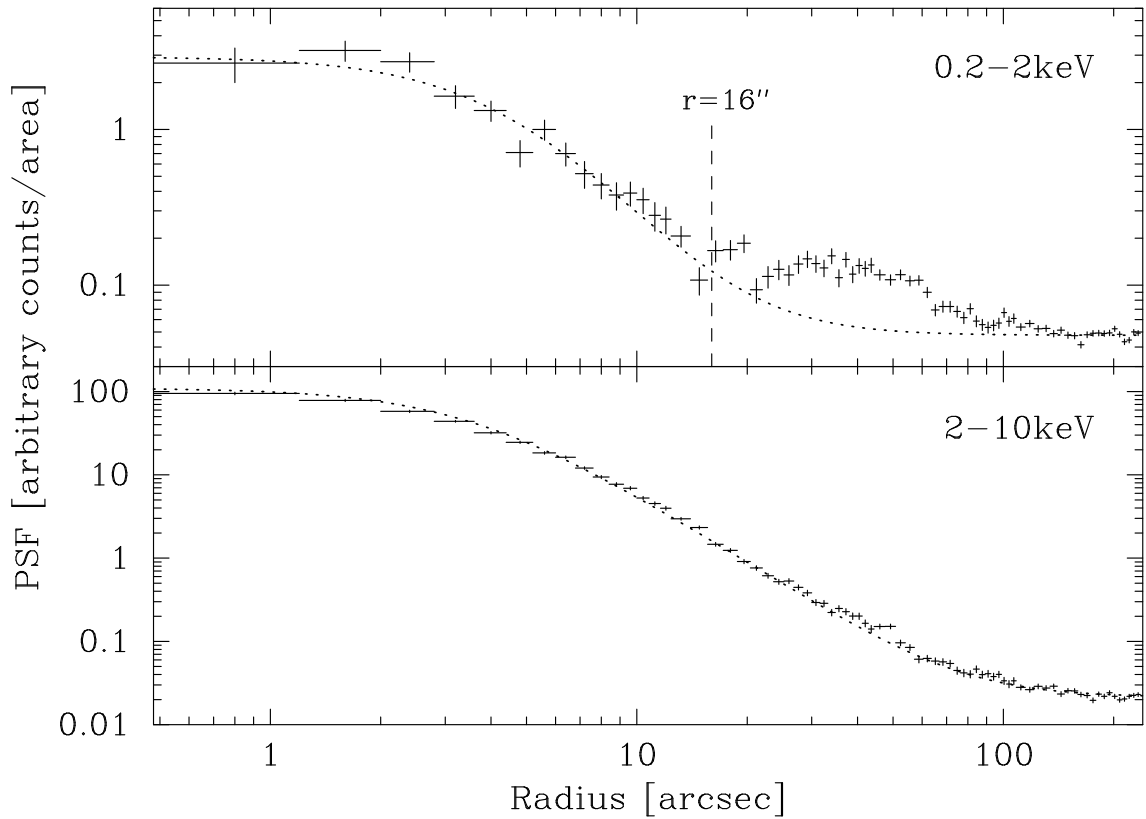}
\caption{
Radial profiles of the nucleus from the MOS detectors in the 
soft band (top) and the hard band (bottom).
The PSF models of King profiles given by \citet{ref:psf}
are shown with constant offset which represents the background.
The hard radial profile follows well along the PSF model; however, 
the soft one shows an excess for $r\sim$~20--60$^{''}$ 
owing to the emission from the host galaxy.
A radius of 16$^{''}$ is adopted for extraction of the soft spectrum.
}
\label{fig:psf}
\end{figure}

We made a radial profile of the nucleus,
in order to investigate whether or not the nuclear emission 
is a point source.
Figure~\ref{fig:psf} shows the radial profile from the MOS detectors
in the soft and hard bands.
The radial profile in the hard band is described well by 
the point-spread function (PSF) model \citep{ref:psf}
plus constant offset, which represents the background. 
Thus, we conclude that the hard emission cannot be distinguished 
from a point source.
However, as suggested by the image, 
the soft radial profile cannot be expressed solely by the PSF model for a point source.
We fitted the data with the King and constant models 
in the regions of $r<10''$ and $r>150''$, respectively. 
In this fit, parameters of the King model other than the normalization
were fixed at the values for a point source given by \citet{ref:psf}.
Although the radial profile shows an apparent excess at
$r\sim$~20--70$^{''}$ that is apparently due to emission in the
galaxy, the core of the observed PSF within $r\la16^{''}$ is
consistent with being a point source,  and therefore the nuclear
emission is unresolved. 

\section{Time Series Analysis}\label{sec:lc}

\begin{figure}
\plotone{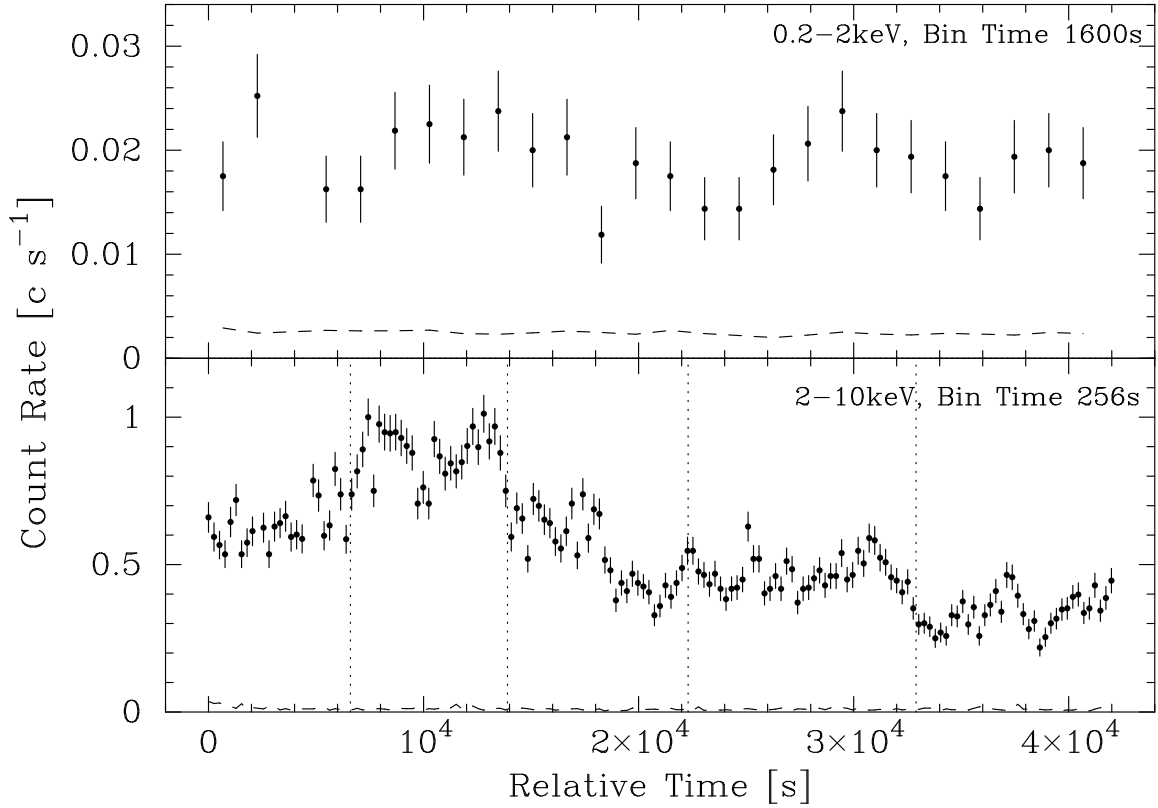}
\caption{The pn light curve of NGC~6300 
in the energy bands of 0.2--2~keV (top) and 2--10~keV band (bottom).
The soft and hard photons 
were collected from the region centered at the X-ray peak
within the radii of 16$^{''}$ and 60$^{''}$, respectively.
The horizontal dashed lines show background levels
(from the BGD1 region).
The vertical dotted lines in the bottom panel
are shown for the time-resolved spectral analyses
in \S\ref{sec:hrdspc}.}
\label{fig:lc}
\end{figure}

Figure~\ref{fig:lc} shows the pn light curves in the soft (0.2--2~keV)
and hard (2--10~keV) bands. 
In the hard band, rapid and rather high amplitude variability 
is clearly seen, whereas the light curve is consistent with constant
in the soft band ($\chi^2=23.5$ for 25 degrees of freedom [dof]).

\begin{figure}
\plotone{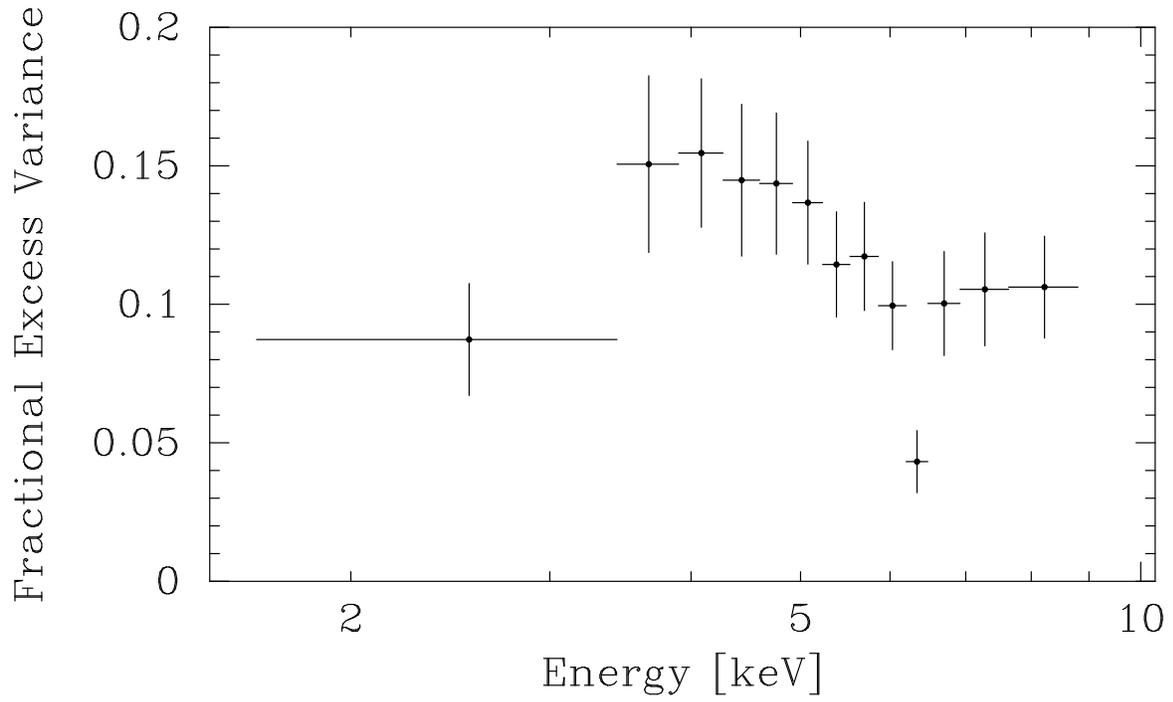}
\caption{Poisson-noise-subtracted variance
normalized by the mean count rate. 
Energy-resolved light curves from the pn and MOS data are binned at 384~s.
Energy bands were chosen so that the same number of photons 
was included in each light curve.
The reduced variability around 6.4~keV implies a less-variable iron line.}
\label{fig:var}
\end{figure}

To search for  spectral variability,
we computed the fractional excess variance as a function of energy 
(Figure~\ref{fig:var}).
Around 2~keV the variance is low; this can be explained by 
dilution of the varying hard component by the less-variable soft emission.  
The larger variance around 4~keV may imply that the spectrum is softer 
when it is brighter.  
The variability drops around 6.4~keV, indicating a less-variable 
iron line.
These points are investigated further in \S\ref{sec:hrdspc}.

\section{Spectral Analysis}

Because of the differences in the radial profiles,
we performed spectral analysis separately in two energy bands.
For the hard-band analysis, 
we extracted the spectra from a circular region with $r<60''$.
The background spectra were collected from a source-free region near 
the nuclear-spectra-extracted region.
For the soft X-ray analysis, we extracted spectra from $r<16''$ regions, 
to avoid contamination from the surrounding galaxy emission.
Each spectrum is grouped so that each energy bin 
has at least 25~photons and so that the energy bin widths are 
about half the detector resolution  ($\sigma\sim35$--70~eV). 
The response matrices were created  using {\sf rmfgen} and {\sf arfgen} 
in {\sf SAS} 5.3.3.
The $\chi^2$ fitting statistics were used.

\subsection{Hard X-ray Spectrum} \label{sec:hrdspc}

\begin{figure}
\plotone{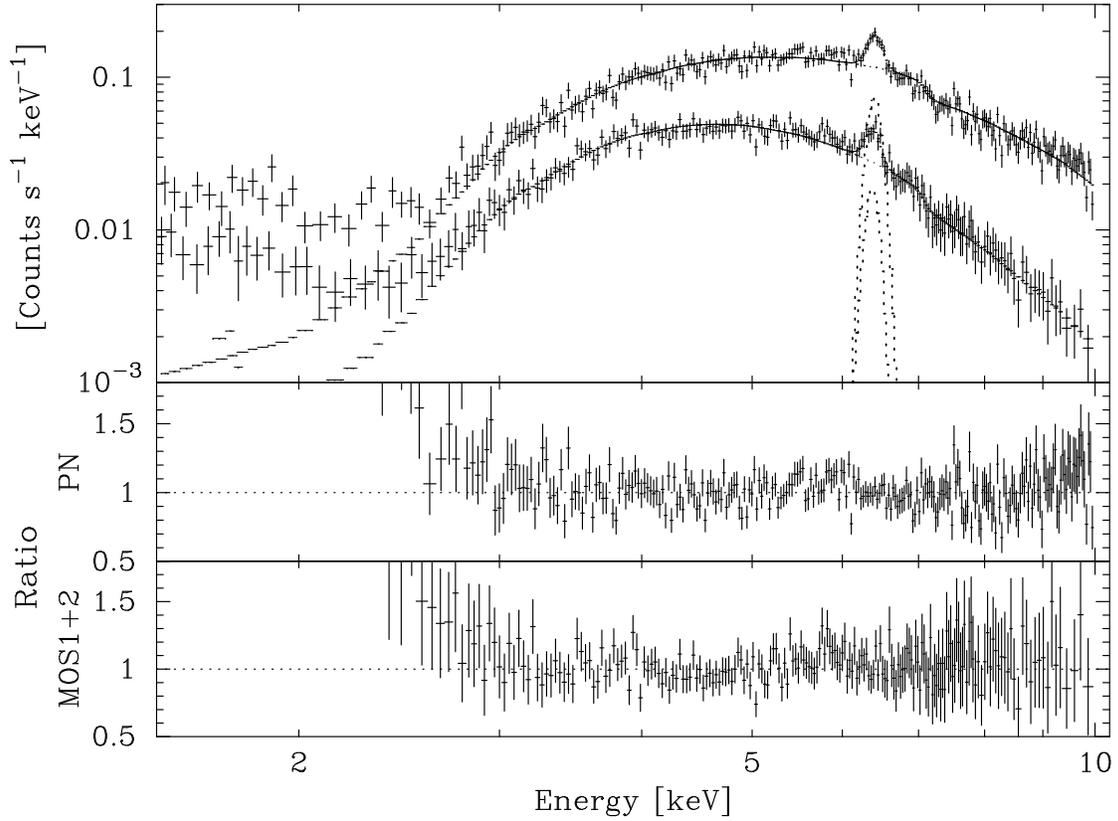}
\caption{Hard X-ray spectra of the nucleus. 
Top: pn and MOS1$+$2 data in the hard X-ray band 
with the best-fit model 
(Compton-thin absorbed power-law plus a Gaussian line).
The model was fitted only in the 3--10~keV band 
and then was extrapolated for the plots.
The middle and bottom panels show the ratio of the data to the best-fit
model for the pn and MOS1$+$2, respectively. These show that the soft
component emerges below $\sim$2.5~keV.
}
\label{fig:hardspc}
\end{figure}

The hard X-ray spectrum of the nucleus peaks at around 5~keV 
(Figure~\ref{fig:hardspc}), 
and it is apparent that intervening column density 
is larger than the Galactic column density  of 
9.38$\times$10$^{20}$~cm$^{-2}$ \citep{GalNh}.
Thus, we fitted the 3--10~keV data with a model of a 
power-law attenuated by Galactic and intrinsic absorption.
The background flux is only a few percent of the source flux
in the hard band. 
As a result of a prominent emission line at $\sim$6~keV, 
the presence of which has been suggested by Figure~\ref{fig:var},
the fits are not acceptable for any EPIC spectra
($\chi_\nu^2=1.96$, 1.41, and 1.30 for 234, 156, and 155 dof
for the pn, MOS1, and MOS2, respectively).
Therefore, we added a Gaussian to the model to represent the line, 
and the fits are improved significantly.
The results are listed in Table~\ref{tab:spc};
the  parameters are consistent among the three detectors.
The fits are statistically acceptable for  both of the MOS spectra; 
for the pn spectrum,  the fit is not acceptable at 90\%
confidence level, however 
it is not rejected at the 99.6\% confidence level 
and  would  be acceptable at 90\%
if there were a 4\% uncertainty in the calibration. 

Next, we performed a simultaneous fit of pn and MOS1$+$2 spectra
with this model
and found that the 3--10~keV spectra are well reproduced with 
a rather typical model for Compton-thin Seyfert 2 galaxies: an absorbed
power-law and a relatively narrow iron line (Figure~\ref{fig:hardspc}).
The photon index ($\Gamma$) is 1.83$\pm$0.08, 
and the intrinsic column density (\Nh) is 
$\sim$2.2$\times10^{23}$~cm$^{-2}$.
The absorption-corrected 2--10~keV luminosity is 
1.3$\times$10$^{42}$~erg~s$^{-1}$, 
assuming $H_0=50$~km~s$^{-1}$~Mpc$^{-1}$.\@\footnote{
The luminosity is 6.2$\times$10$^{41}$~ergs~s$^{-1}$
using  the {\it WMAP} cosmological parameter of $H_0=72$~km~s$^{-1}$~Mpc$^{-1}$.
}
For the line, we obtained the central energy of 6.43$_{-0.02}^{+0.01}$~keV, 
indicating that the iron is not significantly ionized.
The line width $\sigma$ is marginally resolved to be 
$55_{-21}^{+19}$~eV.
If we assume that the line is also attenuated by the intrinsic 
absorber,
the equivalent width (EW) of the iron line is 148$\pm18$~eV.

We also made spectral fits to time-resolved spectra
in order to confirm the suggested variability behavior 
of each spectral component discussed in \S\ref{sec:lc}. 
The light curve was split into five segments such that
each segment corresponds to a particular state of the flux
(i.e., high, low, decreasing, etc.) and so that
the duration of each segment is approximately equal 
to 1$\times$10$^{4}$~s (Figure~\ref{fig:lc}).
We fitted the resulting five sets of spectra in the same manner 
as described in the previous paragraph.
The soft spectra were not examined 
because the photon statistics are poor
and because 
the light-curve analysis revealed 
no significant variability in the soft band.
We performed model fits in which all parameters were allowed to vary;
we found significant variability in only the power-law normalization.
It is noteworthy that the intensity
of the iron line remains constant, despite the factor of 2.6 variation of the
ionizing continuum flux. If the iron line would have instantaneously
responded to the continuum, its variability should have been clearly
detected above the rather small statistical uncertainties 
on the iron line intensity ($\sim$15\%).

The reflection component, which is characterized by a hump
 around 20~keV, 
has been intensively studied using the previous observations 
that were made by satellites with good efficiency above 20~keV.
In the \xmm\  bandpass, which extends to $\sim$10~keV only, 
the reflection component 
is not prominent; thus, the data 
are not very sensitive to 
the reflection model. Therefore, we investigated this model 
holding some spectral parameters fixed.
We made fits using a reflection model from a neutral disk
({\sf pexrav}), assuming that the incident photons on the disk
have a power-law spectrum with the observed power-law index and that
the abundances of the disk are solar. 
We tested the cases in which the cut-off energy of the power-law 
is 100 or 250~keV and the absorption column density for 
the reflection component is only Galactic or is the same as the nucleus.
The choice of the assumed values did not affect the result of the fit
significantly;  
addition of the reflection improved the fits by $\Delta\chi^2\sim$7--10,
and the photon index became slightly steeper ($\Delta\Gamma\sim0.10$).
The parameters of the iron line were not significantly affected.
The contribution of the reflection component is
12\%--17\% of the total model flux in the 3--10~keV band;
in terms of the reflection parameter $R$,\footnote{
$R=1$ corresponds to an illuminated solid angle of 2$\pi$ 
for isotropic incident emission.} 
the flux is equivalent to $R=1.3^{+1.1}_{-0.9}, 1.7^{+1.4}_{-1.1}$, 
and $4.8^{+3.8}_{-3.1}$,  
assuming that we observe through an absorption 
with the same column density as that of the nucleus
at the inclination angle\footnote{These angles are measured from axis of symmetry.}
of 30$^\circ$, 60$^\circ$, and 85$^\circ$, respectively.

\subsection{Soft X-ray Spectrum}\label{sec:sft}

\subsubsection{Background}

The background (BGD) subtraction is 
a difficult issue for the
soft X-ray analysis because 
the soft emission from the active galactic nucleus (AGN) is faint
and because it is not known  a priori
whether the soft X-ray emission 
from the host galaxy extends close to the AGN.
Thus, we considered two BGD spectra sets.

Assuming that the host galaxy emission is confined to the galactic 
ring and 
does not contaminate the nuclear spectrum,
we extracted one set of BGD spectra (BGD1) from the source-free-region 
at the distance $r>2'$. 
We also considered the complementary assumption, that the soft X-ray emission
from the galaxy is approximately uniform through the region. For this, another set
(BGD2) is collected from the region where extended 
emission is prominent, including a rectangular region ($70''\times40''$; 
shown in Figure\ref{fig:img})  and 
excluding the region of $r<20''$ from the nucleus.  
In general, using BGD2 results in somewhat flatter spectral shape 
(e.g., $\Delta\Gamma\sim0.3$) and  $\sim$35\% reduction in flux.

\subsubsection{Absorption of the Reprocessed Emission} \label{sec:RefAbs}

Another difficulty in the soft X-ray spectral analysis is
the question of the magnitude of the absorption of
the reflection component.
If the reflection is attenuated by the same absorption as
the nucleus, the emission is negligible below about 2~keV.
However, since NGC~6300 was seen with \xte\ to be 
reflection-dominated, the reflection may not be attenuated by 
the same absorption as the nucleus. 
In this case, reflection can dominate in the 1--3~keV band.

In order to estimate the magnitude of the absorption 
for the reprocessed emission,
we fitted the 2--10~keV data 
assuming that the reprocessed 
emission is covered by an absorption column density \Nh$_{repr}$, 
which is left free in the fit. 
Although the fit could not constrain \Nh$_{repr}$ 
owing to limited data quality, the best-fit value was  
$\sim2\times10^{22}$~cm$^{-2}$, regardless 
of the assumed reflection parameters and the choice of the background.
If \Nh$_{repr}\sim2\times10^{22}$~cm$^{-2}$, then
the reflection component is insignificant below 1.6~keV.

\subsubsection{Spectral Modeling}\label{sec:sftmdl}

With the absorption estimated in the previous section,
we performed the following fits in the 0.2--1.6~keV band.

\begin{figure}
\plotone{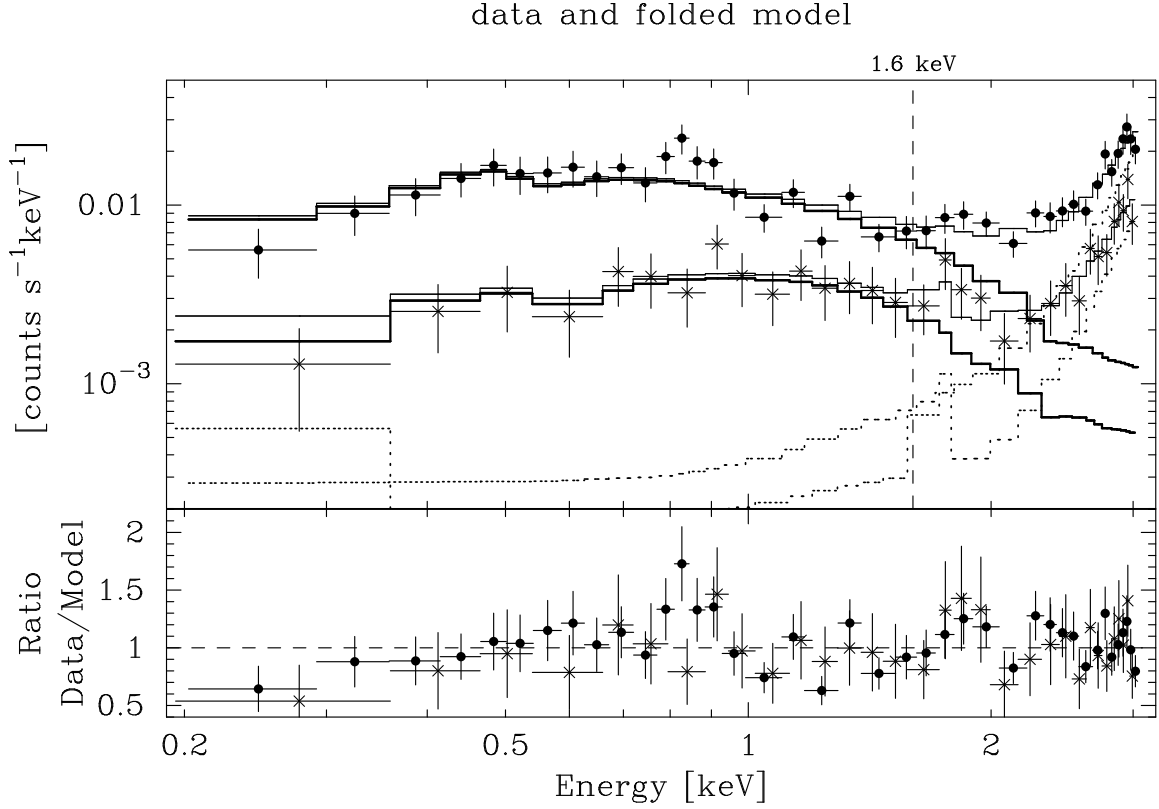}
\caption{Soft X-ray spectra of the nucleus.
The data from the pn and the MOS are marked by filled circles and 
crosses, respectively.
The soft emission was modeled with the Galactic-absorbed power-law (thick solid lines).
BGD1 is used for this figure. 
The fit was performed only in the 0.2--1.6~keV band, to avoid
contamination from the hard X-ray emission (shown by the dotted lines for reference; 
the model parameters were taken from Table~\ref{tab:wholespc}).
The bottom panel shows the ratio of the data to the model including 
the hard components.
The overall spectrum is reproduced by the model, but 
there are hints of emission line-like features around 0.8 and 1.9~keV.}
\label{fig:softspc}
\end{figure}

First, we modeled the soft spectra with a Galactic absorbed power-law
and obtained acceptable $\chi_\nu^2$ values 
(0.92 and 0.62 for 35 dof with BGD1 and BGD2, respectively). 
The spectra and
the best-fit model with BGD1 are shown in Figure~\ref{fig:softspc},
and the parameters are summarized in Table~\ref{tab:sftspc}.
The soft photon index  ($\Gamma_{\rm soft}$) is 
2.00$_{-0.18}^{+0.16}$ and 1.72$_{-0.28}^{+0.26}$ 
using BGD1 and BGD2, respectively.
These values are consistent with the hard photon index 
($\Gamma_{\rm hard}=1.83\pm$0.08), suggesting that
the soft emission could be nuclear continuum scattered by electrons.
The ratio of the soft to 
hard power-law flux is 0.2\%--0.3\% in the 0.2--2.0~keV band.

It is also possible that the soft X-ray emission originates in 
optically-thin hot gas associated with possible starburst activity.
The Raymond-Smith thermal plasma model in {\sf XSPEC} was used
to test this scenario.
The fit results are also listed in Table~\ref{tab:sftspc}.
This model is not rejected statistically, and 
the size of the plasma\footnote{
Assuming that the plasma is uniform in density, $n$~(cm$^{-3}$),
the size of the plasma is estimated to be 
$\sim10^2 (1/n)^{-2/3}$~pc.}
deduced from the X-ray intensity 
can be consistent with the radial profile of the X-ray image.
However, the obtained abundance was small: 
the 90\% upper limit is 0.01 and 0.1 solar
for BGD1 and BGD2, respectively.
With BGD1, this model was rejected at the 99\% confidence level
in the abundance regime larger than 0.3~solar.
Although we cannot rule out this model, 
this model is not preferred because of the very low abundances.

Finally, we considered intrinsic absorption for the soft X-ray 
emission from the nucleus, by adding an additional absorption component 
in the above models. 
The fit results are also summarized in Table~\ref{tab:sftspc}. 
In most cases, intrinsic absorption is not statistically required. 
Only the power-law model fit with BGD1 improved (96\% by the $F$-test); 
however, the 90\% lower-limit of \Nh\ is as small as 
1$\times$10$^{20}$~cm$^{-2}$.
In every fit, the upper-limit of \Nh\ is 1--2$\times$10$^{21}$~cm$^{-2}$.
This result suggests that the soft X-ray--emitting region is not covered
by a significant amount of absorbing gas intrinsic to NGC~6300.

\subsection{Overall Picture}

\begin{figure}
\plotone{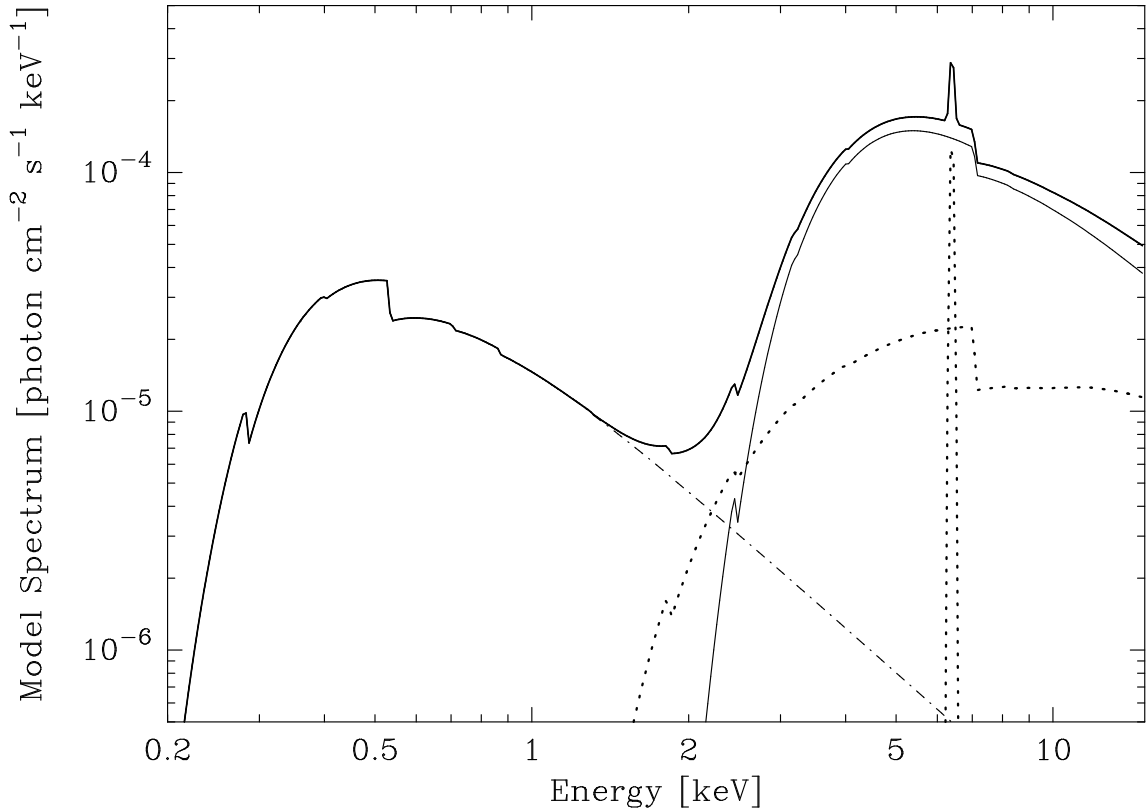}
\caption{Final inferred model spectrum in the whole band. 
The parameters are listed in Table~\ref{tab:wholespc}.
The thick solid  line represents total observed flux.
The thin solid line shows the hard power-law attenuated by the intrinsic absorption of \Nh$_{hard}$. 
The dotted lines represent
the reprocessed emissions (reflection continuum and an iron emission line)
attenuated by another absorber of  \Nh$_{repr}$. 
Although the soft X-ray spectrum is simply modeled by a power-law (dot-dashed line),
the presence of some emission line-like features is suggested 
(see Figure~\ref{fig:softspc} and \S\ref{sec:sft_origin}). 
Thus, the soft emission could be
emission from a  photoionized plasma imprinted 
by many emission lines and radiative recombination edges. 
}
\label{fig:wholespc}
\end{figure}

Finally, we fitted the data in the entire \xmm\ bandpass.
The model included all spectral components that have been
considered before. Specifically, the spectral model consists of 
(1) a hard power-law attenuated by an intrinsic absorption
of \Nh$_{hard}$,
(2) reprocessed emission attenuated by another 
intrinsic absorber of \Nh$_{repr}$,
and (3) a soft power-law without intrinsic absorption.
Assuming that the soft X-ray emission is electron-scattered
nuclear power-law emission and self-absorption is negligible, 
the soft power-law index is constrained to be the same as 
the hard power-law index.
All components are attenuated by Galactic absorption.
The best-fit spectral model is shown in Figure~\ref{fig:wholespc},
and the parameters are summarized in Table~\ref{tab:wholespc}.
This model can reproduce the observed spectrum in the whole
band ($\chi_{\nu}^{2}=0.92$ for 475 dof).

\section{Discussion}

\subsection{On the Origin of the Soft X-ray Emission}
\label{sec:sft_origin}

The \xmm\ observation allowed us to investigate the nuclear
spectrum of NGC~6300 for the first time without 
contamination from the surrounding emission.
We found that it can be modeled
with a power-law having the same photon index as
that in the hard band.
One possible origin for this emission would be that
the soft X-rays are nuclear emission leaking 
through an inhomogeneous absorber.
However, this model is rejected by the smaller variability amplitude 
in the soft band.
Thus, the soft emission can be considered to be nuclear 
emission scattered by electrons with some spatial extent 
that serves to smear out the variability 
of incident nuclear emission. 

Recent high-resolution spectra have revealed that 
soft X-ray emission in some Seyfert 2 galaxies originates from 
photoionized plasma (e.g., Sako et al. 2000; Sambruna et al. 2001).
Unfortunately NGC~6300 is not bright enough to
investigate with the RGS; however, the EPIC spectra suggest
the presence of some emission 
line-like features at about 0.8 and
1.9~keV (see the bottom panel of Figure~\ref{fig:softspc}).  
If we add a narrow ($\sigma=0$) Gaussian to 
the final model (Table~\ref{tab:wholespc}),
the fits improved at the 99\% and 92\% confidence levels (by the $F$-test) with the line 
at 0.84$^{+0.09}_{-0.04}$ and 1.90$^{+0.07}_{-0.12}$~keV, respectively.
These energies suggest that if they are emission lines, 
they may be Fe L- and/or Ne K-lines, as well as Si K-lines; 
if they are radiative recombination continua (RRCs), they may be
those from O and Mg ions.
Regardless of whether they are lines and/or RRCs, the observed energies 
suggest that the ions are highly ionized.
Thus, the plasma should be rich with free electrons, 
and the emission scattered by the electrons can also contribute
to the soft X-ray emission.
Actually, the observed soft spectrum is so smooth that
it cannot be described solely by a model of a photoionized plasma
emission calculated by {\sf xstar2xspec}\footnote{ 
See http://heasarc.gsfc.nasa.gov/docs/software/lheasoft/xstar/xstar.html.} 
because the model predicts more prominent line features if 
abundance is about solar.
In reality, both components are likely to contribute to 
the soft X-ray emission; 
however, because of the limited data quality, 
it is impossible to investigate further.

\subsection{Long-Term Variability and  
Comparison with the Previous Observations}

\subsubsection{Soft X-ray}

\xmm\ observed that the 0.2--2~keV flux from the nucleus is
3--4$\times$10$^{-14}$~erg~cm$^{-2}$~s$^{-1}$
without correcting for Galactic absorption.
Below we compare this value with the previous values from
other instruments.

In 1979, NGC~6300 was observed by the {\it Einstein} IPC
(spatial resolution of $\sim$1$^{''}$), and 
the 3\,$\sigma$ upper limit to the count rate was 
$1.19\times10^{-2}$~counts~s$^{-1}$ between 0.2 and 4.0~keV \citep{fab92}.
This count rate corresponds to the upper limit of the absorption-uncorrected 
0.2--2~keV flux of 
2--4$\times$10$^{-13}$~ergs~cm$^{-2}$~s$^{-1}$,
using the Galactic absorbed power-law model with 
$\Gamma$ in the range of 1--5.

{\it ROSAT} made three observations of NGC~6300 with the HRI
(half power radius is about 4$^{''}$) and detected the object twice.\footnote{
The target was not detected in the shortest observation, which had an
exposure of  only 1~ks.}
The count rates were 
(3.9$\pm$1.3)$\times$10$^{-3}$~counts~s$^{-1}$ in 1997 October
and
(1.1$\pm$0.5)$\times$10$^{-3}$~counts~s$^{-1}$ in 1998 March.\footnote{
From ``ROSAT Source Browser'' at  
http://www.xray.mpe.mpg.de/cgi-bin/rosat/src-browser.}
Although the flux might have decreased between two observations,
the significance of that inference is low.
We converted the weighted mean count rate of 
1.5$\times$10$^{-3}$~counts~s$^{-1}$ into flux using the Galactic absorbed
power-law model with $\Gamma$ in the range of 1--5; 
the count rate corresponds to the absorption uncorrected 0.2--2~keV flux of 
5--6$\times$10$^{-14}$~erg~cm$^{-2}$~s$^{-1}$.

The \sax\ LECS  (half-power radius is 3.5~arcmin at 0.25~keV)  
observed in 1999 that the 0.1--2~keV count rate is $\sim$0.01~counts~s$^{-1}$ 
integrated over the $r<8'$ region from the nucleus \citep{gua02}.
For the best-fit power-law model (Galactic absorbed power-law with 
$\Gamma\simeq4.5$),
the LECS count rate 
corresponds to the 0.2--2~keV flux of 
$\sim$1$\times$10$^{-12}$~ergs~cm$^{-2}$~s$^{-1}$
uncorrected  for Galactic absorption.
However, the integration area is pretty large because of
the \sax\ spatial resolution.
If we look at the 0.2--2~keV \xmm\ image, we can see that 
there were a handful of point-like sources and ``diffuse'' galaxy
emission in the $r<8'$ region.
The fraction of the count rate of the nucleus 
was about 10\% among the total count rate from the r$<8'$ region.
Although the fraction can be time-variable, 
the \sax\ flux should be heavily contaminated with surrounding emissions.
Thus, we regard the soft flux from the \sax\ observation as an upper-limit.

The observed \xmm\ 0.2--2~keV flux of 3--4$\times$10$^{-14}$~ergs~cm$^{-2}$~s$^{-1}$
is consistent with the {\it ROSAT} flux taking into account the uncertainties, and 
is smaller than the upper-limit from the {\it Einstein}  and \sax\ observations. 
Thus, the soft X-ray flux is consistent with being non-variable 
on long time scales, and this result is consistent
with a view that the soft X-ray emission results 
from plasma with large spatial extent.

\subsubsection{Hard X-ray Continuum}\label{sec:LTcont}

Before this \xmm\ observation, 
\xte\ and \sax\ had performed hard X-ray observations of NGC~6300. 
In the hard X-ray band the nuclear emission is dominant 
and is brighter than the second-brightest source in the \xmm\ field
by $\sim$2 orders of magnitude.
Thus, the results from \xte\ and \sax\ are considered to 
be free of contamination from serendipitous emission.

The hard X-ray fluxes from the multiple missions 
are summarized in Table~\ref{tab:longterm}.
The observed 2--10~keV flux from this \xmm\ observation
is 8.6$\times$10$^{-12}$~ergs~cm$^{-2}$~s$^{-1}$,
which is between the \sax\ flux of 1.3$\times$10$^{-11}$~ergs~cm$^{-2}$~s$^{-1}$
and the \xte\ flux of 6.4$\times$10$^{-12}$~ergs~cm$^{-2}$~s$^{-1}$\@. 
The column density of the Compton-thin absorber of the hard power-law 
is consistent with a constant between the \xmm\ and \sax\ observations.

The high ratio of the \xte\  flux to the \xmm\ flux (74\%)
indicates that the bright reflection component  
seen by \xte\ 
was not present during the \xmm\ observation.
In fact, the \xte\  Compton reflection model predicts higher
flux in the 1--3~keV energy band than was observed in the \xmm\
spectra. In order to suppress the model-predicted flux 
so that it is smaller than
the observed one using neutral absorption, a column density larger than 
5$\times$10$^{22}$~cm$^{-2}$ is required. 
If such absorption were present,
the \xte\ flux from reflection 
in the 2--10~keV band  would be 
5.1$\times$10$^{-12}$~erg~cm$^{-2}$~s$^{-1}$, 
which would still comprise 60\% of the observed \xmm\ 2--10~keV flux.
Can the \xmm\ spectrum accommodate such large reflection?
To test this idea,  we fitted the 2--10~keV \xmm\ spectra
with a model consisting of this reflection, an absorbed power-law, and an iron line.
This fit yields $\Gamma=2.6\pm0.2$, which is notably steep among Seyfert 1 and 2 galaxies.
If we allow the normalization to vary while retaining the shape of 
the reflection component,
we find that the normalization of the reflection should be only 
26$^{+25}_{-15}$~\% of that observed by \xte.
Thus, it seems that the flux of the reflection component
has decreased over the 4~yr 
spanned by the \xte\ and \xmm\ observations.
Since these observations are separated by 4 yr, 
such long-timescale variability would be consistent with reflection
from the putative torus, which is considered to be located at about 1~pc
from the central engine.

\citet{gua02} has already reported that the Compton reprocessing matter
should be located within $\simeq$0.75~pc 
based on the large value of the relative Compton-reflection flux 
($R=4.2^{+2.6}_{-1.7}$) measured during the \sax\ observation. 
He argues that it can be explained by considering that 
(1) the AGN was switched off during the \xte\ observation, 
(2) it was switched on between the \xte\ and \sax\ observations 
and was brighter than during the \sax\ observation, 
and (3) the reprocessed emission during the \sax\ observation was
echoing the past glorious state. 
However, if the uncertainties are taken into account, 
the change of the reprocessed continuum flux 
between \xte\ and \sax\ observations
is statistically marginal.\footnote{
By comparing the fluxes at 20~keV of Figure 5 in his paper, 
the variation should be less than a factor of 2. 
Even at 20~keV, the power-law flux still contributes 
$\sim$25\% of the total flux in the \sax\ spectrum. 
The uncertainty of reflection normalization $R$ is as large as 
$\sim$50\% according to his Table~1.}
Thus, considering only the \xte\ and \sax\
results, we cannot distinguish between the possibilities that 
the reprocessing material is located farther away from the central
source and that the reflection component remains constant.  We infer
from the  \xmm\ observation that a significant reduction in the
reprocessed flux has occurred, so we can now rule out the possibility
that the reflection flux is constant.

\subsubsection{Iron Emission Line}

The intensity of the iron emission line from \xmm\ 
shows significant reduction in flux, compared with that from \xte\
(Table~\ref{tab:longterm}).
The intensity inferred from the \sax\  spectrum is between those
obtained from the \xte\ and \xmm\ spectra,
although the \sax\ value is consistent with either of the \xte\ and
\xmm\ values, within the uncertainties. 
Since the iron-line flux and the reflection flux
are considered to be  time-averaged echoes of nuclear emission,
the observed trend of decreasing reprocessed flux suggests 
that the nucleus has experienced an overall long-term trend 
of decreasing intensity on a timescale of years.

It should be noted that
the EW of the iron line with respect to the reflection continuum
is consistent with being constant 
among the three observations. This constancy is expected, if the iron
line and reflection originate in the same gas and hence
reflect the past nuclear activity at the same period.

The iron-line EW with respect to the reflection continuum
was found to be 400--500~eV, combining the results from 
the three hard X-ray observations.
The EW with respect to the reflection continuum is determined 
by geometry and abundance.
For a disk geometry, it is calculated to be 1--2~keV 
(depending on the inclination angle; solar abundance is assumed;
 Matt et al., 1991).  
The observed EW is smaller than the theoretical expectation
assuming solar abundance of iron. 
For a torus geometry, the predicted EW is also about 1~keV 
(e.g., Figure~3 of Matt et al., 2003), 
and thus the observed EW is again somewhat smaller.
This result may suggest that iron abundance is subsolar,
as has already been pointed out by \citet{lei99}.

\subsection{On the Origin of the Iron Emission Line}
\label{sec:FeOrigin}

The \xmm\ observation allowed us to examine the iron K-$\alpha$ emission
line with sufficient energy resolution to distinguish its
ionization level. We found from the line energy 
that the iron is not significantly ionized.
The line width is marginally resolved to be 
$\sigma\sim55^{+19}_{-21}$~eV;
the detector resolution is $\sigma\sim60$~eV at 6.4~keV.
To examine this width with the best possible energy resolution,
we created spectra only from the single-pixel (pattern$=$0) events. 
The fits to the spectra from the pn, MOS1, and MOS2
yield the line widths of
56$^{+25}_{-31}$, 100$^{+59}_{-49}$, and 
59$^{+62}_{-59}$~eV, respectively.
Thus, the line profile in each detector favors $\sigma\sim$60~eV independently.
If we assume that this line width originates from gas with 
Keplerian velocity,
the line-emitting region is estimated 
to be located at approximately
10$^4$~Schwarzschild radii.
However, the observed line width could instead be a result of
a blend of ionization states.
Then, the iron line emitter could be located at radii larger than
$\sim$10$^4$~Schwarzschild radii. 

Variability is also a useful tool in constraining
the location of the line-emitting region.
Lack of  short-term variability during the 
\xmm\  observation (the duration is $\sim$5$\times$10$^4$~s) 
implies that the region is separated from the nucleus by
$\ga10^{-4}$~pc in order to smear out the variability of the
continuum.  
On the other hand, long-term variability of the iron line was observed.
>From the long-term variability,  
the region is inferred to be located within $\la$1~pc from the nucleus. 
The inferred distance 
suggests that
the region reprocessing the iron line is 
the outer part of the accretion disk and/or the torus.

\subsection{Short-Term Variability}

It should be noted that we detected rapid and rather high amplitude variability
(Figure~\ref{fig:lc});
such variability is rarely detected
among Seyfert 2 galaxies (e.g., Turner et al. 1997).
\citet{ref:Sy2ASCA}  analyzed {\it ASCA} data from 25 Seyfert 2 galaxies 
(including eight narrow emission line galaxies).
Of these, only five objects were bright enough 
to investigate short-term variability,
and three of the five objects\footnote{All three are 
narrow emission line galaxies, and 
the only Seyfert 2 galaxy (NGC~1068) that 
was bright enough to investigate 
short-term variability did not show significant 
short-term variability.}
were found to lie within the trend from a Seyfert 1 sample
\citep{ref:Sy1ASCA}.
To compare our NGC~6300 result with their Seyfert 1 and 2 samples, 
we quantified the variability in the same manner 
as they do
using the normalized excess variance ($\sigma^2_{\rm RMS}$) 
for the light curve binned at 128~s in the 0.5--10~keV band.
The value of $\sigma^2_{\rm RMS}$=0.107$\pm$0.009 from the pn
and the 2--10~keV intrinsic luminosity of 
1.2$\times$10$^{42}$~ergs~s$^{-1}$ 
are consistent with the trend from their Seyfert 1 sample.
\citet{awa04} provide more discussion on variability properties 
  of NGC 6300 in comparison with other Seyfert galaxies and estimates 
  of central mass using the power density spectrum, as well as other methods.

\subsection{On the Optical Reddening}

The soft X-ray spectrum suggests that the intrinsic absorption of 
this system
is less than \Nh$=$1--2$\times$10$^{21}$~cm$^{-2}$ (\S\ref{sec:sftmdl}); 
however, the optical spectrum from NGC~6300 is highly reddened 
(e.g., Lumsden, Alexander, \& Hough 2004).  
Based on the fact that the emission lines
are polarized to the same degree as the continuum (e.g., Lumsden et
al.\ 2004) and on the presence of a prominent dust-lane that may lie in
the line of sight to the nucleus (e.g., Maddox et al.\ in prep.), it
seems plausible that the polarization and reddening originate in the
host galaxy.  If so, there should be gas associated with the dust, and
that gas should attenuate the X-ray spectrum.

We can estimate the intrinsic column as follows.  \citet{lum04}
report an observed Balmer decrement (H$\alpha$/H$\beta$) of 8.7.  We
note that this value is uncertain. The optical nucleus of NGC~6300 is
weak, and any observation is likely to contain a significant amount of
light from the host galaxy; the presence of stellar spectral features
(e.g., \ion{Na}{1}~D) can be seen in the spectrum published by \citet{lum04}.
Thus, the nuclear continuum and the H$\beta$ emission line may be
difficult to distinguish from the galaxy light and stellar H$\beta$
absorption line without careful template subtraction (e.g., Halpern \&
Fillipenko 1994).  Regardless, using the Cardelli, Clayton, and Mathis
(1989) reddening law, we infer that the Balmer decrement would be 7.9
after accounting for the reddening in our Galaxy [E(B$-$V)=0.097;
Schlegel et al.\ 1998].  If the intrinsic H$\alpha$/H$\beta$ ratio is
3.1 (Halpern \& Steiner 1983), a Balmer decrement of 7.9 corresponds
to E(B$-$V) of 0.89. This corresponds to $N_{H} \approx 3.7 \times
10^{21}\rm\,cm^{-2}$, if the dust-to-gas ratio in NGC 6300 is the same
as it is in our Galaxy (Heiles, Kulkarni \& Stark 1981).

The deduced column density of the absorbing gas is much lower 
than that deduced from the hard X-ray spectral shape, 
as Maiolino et al. (2001) found a quite low E(B$-$V)/$N_{H}$ among 
Seyfert galaxies. However, 
our estimate of E(B$-$V) is based on extinction for narrow lines;
hence, it should be rather distant from the nucleus.
Therefore, the dust may not be cospatial with the gas absorbing 
the hard X-ray, as we discussed in \S\ref{sec:RefAbs}
regarding this object, 
and as Weingartner \& Murray (2002) suggested in general.

On the other hand, the absorption of the soft X-ray emission 
could cospatially exist with the dust optical extinction
because soft X-ray emissions from photoionized plasma are
reported to be associated with the (extended) narrow-line region
(e.g., Sako et al. 2000; Yang,  Wilson \& Ferruitet 2001).
We fitted the 0.2--1.6~keV spectra with a model including
intrinsic absorption while \Nh\ was fixed at the value
deduced above.
The fits with a power-law model were not rejected statistically
($\chi_{\nu}^2=1.17$ and 0.84 for 34~dof
for BGD1 and BGD2, respectively).
However, the  power-law index obtained is quite steep
($\Gamma=4.8\pm0.2$ and $4.3\pm0.4$ for BGD1 and BGD2, respectively),
suggesting that this model is unphysical
by contrast with an origin of the power-law as energy-independent
electron scattering of the primary continuum.
Thus, if the soft X-ray emission is seen through the 
same medium as optical narrow emission-lines,
the dust condition in the medium may be different from that 
in our Galaxy;
for example, the deduced smaller \Nh/E(B$-$V) suggests that
NGC~6300 may have a larger dust-to-gas-ratio and/or 
smaller grain size.

\section{Summary}

We investigated the nucleus of NGC~6300 using the results
from a half-day observation by \xmm.
The X-ray spectrum of the nucleus 
consists of a heavily absorbed
hard component dominating in the 3--10~keV band 
and a soft component seen in the 0.2--2~keV band.  In the hard
band, the spectrum is well fitted by a power-law model
with a photon index of 1.83$\pm$0.08 
attenuated by a Compton-thin absorber
($N_{\rm H}\simeq 2.2 \times 10^{23}\,\rm cm^{-2}$). 
A narrow iron line is detected at 6.43$_{-0.02}^{+0.01}$~keV with EW$\sim$150~eV;
the line velocity width is marginally resolved to be $\sigma\sim60$~eV.
The observed properties of the iron line suggest that
the origin is the outer part of the accretion  disk and/or  the
putative torus. 
The decline of both the iron line and the reflection continuum fluxes 
on a time scale of years is observed, which implies that the nucleus
has experienced an overall long-term trend 
of decreasing intensity on the same timescale.
The Compton-thin absorber attenuating the hard X-ray continuum
is more compact than a few hundred parsecs. 
Soft X-ray emission is likely to be electron-scattered light
from a spatially extended region, and emission from a photoionized plasma
is also likely to contribute to it.

\acknowledgments

This work is based on the observation obtained with \xmm,
an ESA science mission with instruments and contributions 
directly funded by ESA Member States and NASA.
We acknowledge the great efforts of 
all of the members of the \xmm\  team.
We thank Dr. Guainazzi for providing 
a complete set of parameters for his \sax\ spectral model
and an anonymous referee for useful comments.
K.M.L. acknowledges useful conversations with Rick Pogge regarding
nuclear reddening.
This work is partially supported by the \xmm\ AO1 grant (NAG 5-9991).

\clearpage

\clearpage

\clearpage

\clearpage 

\clearpage 

\clearpage

\clearpage

\clearpage

\begin{deluxetable}{cccccccc}
\tablecaption{
Results from the fits in the 3--10~keV band
 \label{tab:spc}}
\tablehead{
\colhead{Data} 
& \colhead{\Nh\tablenotemark{a}}  
& \colhead{$\Gamma$}   
& \colhead{$F_{\rm 2-10keV}$\tablenotemark{b}}  
& \colhead{$E_{\rm Fe}$~[keV]}  
& \colhead{$\sigma_{\rm Fe}$~[eV]}  
& \colhead{$F_{\rm Fe}$\tablenotemark{c}}  
& \colhead{$\chi^2$/dof}   
}
\startdata
pn & $2.14_{-0.10}^{+0.10}$	& 1.85$_{-0.09}^{+0.09}$	
	& 2.16$_{-0.12}^{+0.11}$	& 6.43$_{-0.02}^{+0.01}$
	& $49_{-32}^{+24}$	& 3.1$_{-0.4}^{+0.5}$	& 291.7/231\\
M1 & $2.03_{-0.16}^{+0.17}$	& 1.60$_{-0.18}^{+0.18}$	
	& 2.05$_{-0.16}^{+0.18}$	& 6.42$_{-0.02}^{+0.03}$
	& $62_{-45}^{+37}$	& 3.5$_{-0.9}^{+0.9}$	& 164.7/153\\
M2 & $2.08_{-0.17}^{+0.17}$	& 1.81$_{-0.18}^{+0.19}$	
	& 2.05$_{-0.17}^{+0.20}$	& 6.40$_{-0.03}^{+0.02}$
	& $55 (<38)$	& 3.0$_{-0.7}^{+0.9}$	& 151.5/152\\
\hline
pn\&MOS
	& 2.15$_{-0.09}^{+0.08}$ & 1.83$_{-0.08}^{+0.08}$	
	& 2.16$_{-0.10}^{+0.10}$	& 6.43$_{-0.02}^{+0.01}$
	& $55_{-21}^{+19}$	& 3.2$_{-0.4}^{+0.4}$	& 423.5/420\\
\enddata
\tablenotetext{a}{Hydrogen equivalent column density of intrinsic absorption
in the unit of $10^{23}$~cm$^{-2}$. }
\tablenotetext{b}{Absorption-corrected power-law flux in the 2--10~keV band. The units are 10$^{-11}$~erg~cm$^{-2}$~s$^{-1}$.}
\tablenotetext{c}{Iron line flux in the units of $10^{-5}$~photon~cm$^{-2}$~s$^{-1}$.}
\tablecomments{
Here we assumed that the line is also attenuated by the same intrinsic absorption
as the continuum; if the line is observed without intrinsic absorption,
the line intensity becomes weaker (transmission efficiency of 
\Nh$\sim2.2\times10^{23}$~cm$^{-2}$ absorption is 69\% at 6.4~keV)
}
\end{deluxetable}

\clearpage

\begin{deluxetable}{ccccccc}
\tablecaption{
The soft X-ray band fits  
with a power-law (PL) or an optically-thin thermal model (RS)
 \label{tab:sftspc}}
\tablewidth{0pt}
\tablehead{
\colhead{Model} 
& \colhead{BGD}  
& \colhead{Intrinsic Abs.\tablenotemark{a}}  
& \colhead{$\Gamma$ or $kT$~[keV]}   
& \colhead{$F_{\rm 0.2-2keV}$\tablenotemark{b}}  
& \colhead{$Z\,$\tablenotemark{c}}  
& \colhead{$\chi^2$/dof}   
}
\startdata
PL & BGD1 	
	 & \nodata &  2.00$_{-0.18}^{+0.16}$	
	 & 7.2 (3.7)	& \nodata & 32.36/35\\
   & BGD2	
	 & \nodata	&  1.72$_{-0.28}^{+0.26}$	
	 & 4.9 (2.8)	& \nodata & 21.80/35\\
  & BGD1 	
	 & $0.9^{+1.1}_{-0.8}$	&  2.8$_{-0.7}^{+0.8}$	
	 & 15. (3.5)	& \nodata & 28.47/34\\
   & BGD2	
	 & 0.5 ($<$1.5)	&  2.1$_{-0.9}^{+1.2}$	
	 & 6.7 (2.7)	& \nodata & 21.34/34\\ \hline
RS & BGD1
	 & \nodata & 1.2$_{-0.2}^{+0.4}$	
	 & 6.6 (3.6)	& 0 ($<$0.01) &29.99/34\\
   & BGD2
	 & \nodata	& 2.1$_{-0.8}^{+3.2}$	
	 & 4.6 (2.8)	& 0 ($<$0.14) &21.57/34\\
   & BGD1
	 & 0.2 ($<$0.8) & 1.0$_{-0.4}^{+0.5}$	
	 & 7.5 (3.5)	& 0 ($<$0.01) &29.34/33\\
   & BGD2
	 & 0.05\,($<$1.0)	& 2.0$_{-1.2}^{+3.3}$
	 & 4.7 (2.7)	& 0 ($<$0.12) & 21.56/33\\
\enddata
\tablenotetext{a}{Hydrogen equivalent column density of intrinsic absorption
in the units of $10^{21}$~cm$^{-2}$. }
\tablenotetext{b}{
Flux in the 0.2--2~keV band in units of $10^{-14}$~erg~cm$^{-2}$~s$^{-1}$\@.
The fluxes are corrected for absorption;
the values in parentheses are absorption-uncorrected fluxes.
}
\tablenotetext{c}{Inferred abundance in units of solar. }
\end{deluxetable}

\clearpage

\begin{deluxetable}{cccccccccc}
\tabletypesize{\scriptsize}
\tablecaption{The best-fit model in the whole energy band
\label{tab:wholespc}}
\tablewidth{0pt}
\tablehead{\colhead{$\Gamma$} 
& \colhead{Norm$_{soft}$\tablenotemark{a} }
& \colhead{\Nh$_{hard}$\tablenotemark{b} }
& \colhead{Norm$_{hard}$\tablenotemark{a} }
& \colhead{\Nh$_{repr}$\tablenotemark{c} }
& \colhead{$R$\tablenotemark{d} }
& \colhead{$E_{\rm Fe}$ }
& \colhead{$\sigma_{\rm Fe}$}
& \colhead{$F_{\rm Fe}$}
& \colhead{$\chi^2$/dof}
\\
\colhead{ } 
& \colhead{$\times$10$^{-5}$}
& \colhead{[10$^{23}$\,cm$^{-2}$]}
& \colhead{$\times$10$^{-3}$}
& \colhead{[10$^{23}$\,cm$^{-2}$]}
& \colhead{[$\Omega/2\pi$]}
& \colhead{ [keV]}
& \colhead{[eV]}
& \colhead{[photon\,cm$^{-2}$\,s$^{-1}$]}
& \colhead{}
}
\startdata
1.94$\pm$0.09 & 
1.83$^{+0.14}_{-0.13}$ & 
2.4$^{+0.1}_{-0.2}$ &
7.7$^{+1.4}_{-1.2}$ & 
0.4$^{+0.3}_{-0.2}$ &
1.1$^{+1.2}_{-0.6}$ &
6.42$^{+0.02}_{-0.01}$ &
49$^{+21}_{-28}$ &
(2.4$\pm$0.3)$\times$10$^{-5}$  &
438.73/475
\enddata
\tablenotetext{a}{Power-law normalization at 1~keV in the rest frame. The units are
photon~cm$^{-2}$~s$^{-1}$.}
\tablenotetext{b}{
Column density of the absorption for the hard power-law emission.
}
\tablenotetext{c}{
Column density of the absorption for the reprocessed components 
(reflection and a fluorescent iron line).
}
\tablenotetext{d}{Reflection strength. The incident emission is assumed
to be same as the observed hard power-law emission.
The other parameters were taken from Leighly et al. (1999;
inclination angle of 77$^{\circ}$ from the normal, abundances of 0.66 solar, 
power-law cut-off of 500~keV), 
and fixed at those values while fitting.}
\end{deluxetable}

\clearpage

\begin{deluxetable}{clclll}
\tabletypesize{\scriptsize}
\tablecaption{
Flux change in the hard X-ray band during 4~years 
 \label{tab:longterm}}
\tablewidth{0pt}
\tablehead{
& & \colhead{Total\tablenotemark{a}}
& \colhead{Reflection\tablenotemark{b}}
& \colhead{Iron-Line\tablenotemark{b}}
& \colhead{Iron EW\tablenotemark{c}}\\
\multicolumn{2}{c}{Observation} &
 [erg~cm$^{-2}$~s$^{-1}$] & [$10^{-12}$~erg~cm$^{-2}$~s$^{-1}$]
 & [10$^{-5}$~photon~cm$^{-2}$~s$^{-1}$] 
 & ~~~~~~[eV] 
}
\startdata
\xte & 1997 \,Feb 
	& 6.4$\times10^{-12}$	& ~~~~~~~~6.4
& ~~~~~4.7$^{+1.2,+1.2\,e}_{-1.0,-1.2}$ & 470$^{+120,+200\,e}_{-100,-150}$\\ 
\sax & 1999 Aug 
	& 1.3$\times10^{-11}$	& ~~~~~~~~6.5$^{+4.0}_{-2.6}$\,$^d$
& ~~~~~3.6$\pm$1.7 & 310$^{+190}_{-240}\,^f$\\
\xmm & 2001 Mar 
	& 8.6$\times10^{-12}$	& ~~~~~~~~1.3$^{+1.4}_{-0.7}$\,$^d$
& ~~~~~2.4$\pm$0.3 & 990$^{+1030}_{-560}\,^g$\\
\enddata
\tablenotetext{a}{Observed fluxes in the 2--10~keV band.}
\tablenotetext{b}{The fluxes were corrected for absorption.  The reflection fluxes 
are in the 2--10~keV band.}
\tablenotetext{c}{Iron-line EW with respect to the reflection component.}
\tablenotetext{d}{The flux error is estimated by 
propagating the uncertainties in  $R$. In principle, this is also subject to
the uncertainty in the  absorption.}
\tablenotetext{e}{The first one is the statistical error;
the second one is an estimate of the systematic error obtained by
changing the normalization of the background.}
\tablenotetext{f}{The EW error is estimated assuming that
the errors of $R$ and the iron-line intensity
are independent.}
\tablenotetext{g}{Since the line flux is constrained much better than the reflection flux, 
the EW error is estimated by scaling the error in $R$.}
\tablecomments{
The \xmm\ parameters are obtained using  spectral model in Table~\ref{tab:wholespc}.
The parameters for \xte\ and \sax\ are from Table~1 ``Compton Reflection Model'' of \citet{lei99} 
and Table~1 ``Model~3'' of \citet{gua02}, respectively.
The \sax\ model assumed that
both the reflection and the iron-line are attenuated by
the absorption same as the power-law emission
(\Nh\,$=$ 2.2$\times$10$^{23}$~cm$^{-2}$).
Assuming that the iron-line is not attenuated by the absorption,
the flux is about 30\% weaker.
In our \xmm\ fit, we assumed that the reprocessed emission is 
attenuated by an absorber different than that for the nuclear power-law emission. 
The absorption column density of the reprocessed emission is 
4$\times$10$^{22}$~cm$^{-2}$, 
and its attenuation magnitude is about 7\% at 6.4~keV.
}

\end{deluxetable}

\end{document}